\documentclass{article}
\usepackage{spconf,amsmath,graphicx}

\usepackage{enumitem}
\setlist{nosep, leftmargin=14pt}

\usepackage{mwe} 

\title{sMRI-based Brain Age Estimation in MCI using Persistent Homology}
\name{Debanjali Bhattacharya$^{\star}$ \qquad Neelam Sinha$^{\dagger}$}
\address{$^{\star}$ Dept.of Artificial Intelligence, School of Artificial Intelligence, Bengaluru, \\ Amrita Vishwa Vidyapeetham, India  \\
    $^{\dagger}$ Centre for Brain Research, Indian Institute of Science, Bangalore, India \\}

\begin{document}
%
\maketitle

\begin{abstract}
    In this study, we propose the use of persistent homology- specifically Betti curves for brain age prediction and for distinguishing between healthy and pathological aging. The proposed framework is applied to 100 structural MRI scans from the publicly available ADNI dataset. Our results indicate that Betti curve features, particularly those from dimension-1 (connected components) and dimension-2 (1D holes), effectively capture structural brain alterations associated with aging. Furthermore, clinical features are grouped into three categories based on their correlation, or lack thereof, with (i) predicted brain age and (ii) chronological age. The findings demonstrate that this approach successfully differentiates normal from pathological aging and provides a novel framework for understanding how structural brain changes relate to cognitive impairment. The proposed method serves as a foundation for developing potential biomarkers for early detection and monitoring of cognitive decline.
    
\end{abstract}



\section{Introduction}
Brain aging, evidenced by structural brain atrophy over time \cite{ref1,ref2}, significantly influences cognitive decline and signifies the onset/progression of neurodegenerative diseases. As individuals age, alterations in brain morphology—such as cortical thinning, ventricular enlargement and appearance of white matter lesions become increasingly prevalent. However, discrimination between healthy aging-related changes and pathological alterations, such as those associated with Mild Cognitive Impairment (MCI), presents a critical challenge, especially since similar brain regions are often affected in both healthy aging and MCI.
Hence understanding the impact of aging on the brain becomes increasingly important. Such understanding is vital not only for identifying individuals at high risk for cognitive disorders but also for monitoring disease progression and for personalizing interventions. 
One emerging approach is to compare an individual's brain age, estimated from neuroimaging data with their chronological age. Discrepancies between them indicate signs of cognitive decline or heightened the risk of neurodegeneration, even in the absence of clinical symptoms. 
Researchers have identified several qualitative biomarkers that can be used to assess aging based on the physical structure of the brain. For instance, it is well established that there is a strong correlation between the thinning of the cerebral cortex and increasing age. Additionally, aging is often accompanied by the appearance of lesions and general atrophy within the brain. However, despite the numerous qualitative markers available, there is a lack of clear methods to measure these changes quantitatively.
Accurately determining a patient's age based solely on their brain anatomy can be highly beneficial for various reasons. For example, it enables a better understanding of the physical changes associated with normal aging, allowing for the identification of anomalies that may indicate structural changes due to conditions like Alzheimer’s disease. 
Traditional morphometric approaches often fail to capture subtle or global structural changes, particularly in the early stages of pathological aging. Furthermore, these methods typically rely on region-specific metrics, potentially overlooking the complex, holistic changes in brain topology.

While prior studies have extensively investigated functional connectivity and regional volumetric changes in the aging brain using MRI \cite{ref4,ref8,ref9}, few have explored topological alterations in brain structure using advanced mathematical tools like persistent homology \cite{ref3,ref5,ref6,ref7}. Persistent homology is a key method from topological data analysis, offers a powerful framework to study global shape characteristics and multiscale features of brain anatomy. However, its application to brain-age estimation and MCI classification remains underexplored. In this study, we explore changes in brain connectivity by examining alterations in its topology. More specifically, the present study seeks to address the research question:
\textit{Can topological features derived from persistent homology, specifically Betti curves extracted from structural MRI data, be effectively used to (i) predict brain age, (ii) differentiate between normal and pathological aging and (iii) help identify clinical biomarkers that correlate with deviations between predicted and chronological brain age?}
We propose a novel topological framework that leverages Betti curve features from persistent homology to estimate brain age and distinguish MCI from normal aging, facilitating the detection of subtle structural biomarkers indicative of pathological aging.

\begin{figure}
\centerline{\includegraphics[width=8cm]{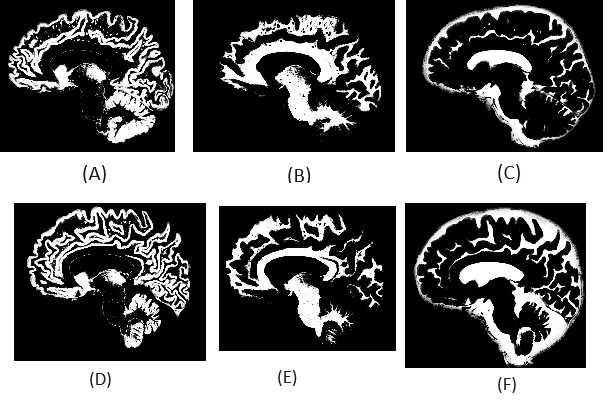}}
\caption{The first, second, and third columns display segmented MRI images of gray matter (GM), white matter (WM), and cerebrospinal fluid (CSF) for a mid-slice of a representative subject. The top row shows images for healthy controls (HC) in Figures (A) GM, (B) WM, and (C) CSF, while the bottom row presents the same for individuals with mild cognitive impairment (MCI) in Figures (D) GM, (E) WM, and (F) CSF.} 
\label{fig1}
\end{figure}

\section{Materials and Methods}

\subsection{Dataset Description}

For this study, T1-W structural 3T MRI scans of ADNI dataset are utilized, which comprises 50 individuals with MCI and 50 healthy controls (HC). In MCI cohort, individuals diagnosed with MCI met specific criteria, including a Clinical Dementia Rating (CDR) score of 0.5, which is currently considered the gold standard for assessing the stages of dementia. The individuals with MCI displayed no additional neurodegenerative conditions. The cognitively healthy group had no history of cognitive decline, stroke, or noteworthy psychiatric ailments. From these sMRI scans,  gray matter (GM), white matter (WM), and cerebrospinal fluid (CSF) segmentation is performed for subsequent analysis. All sMRI scans underwent standard preprocessing steps, including skull-stripping, bias-field correction, intensity normalization, and segmentation into GM/WM/CSF, with quality control ensured through automated checks and visual inspection.
The segmented images for one representative subject of HC and MCI is shown in Figure~\ref{fig1}.

\subsection{Persistent Homology and Betti Curves}
\label{sec:method}
In the proposed approach, persistent homology is employed to examine the topological structure of data across multiple scales or function values, capturing the evolution of topological features. For this, Betti curves can be derived from each $n$-dimensional Betti numbers ($\beta_{n}$), which provide a summary of the topological features present in the data over these varying scales. Betti numbers essentially count the number of topological features like connected components, loops, and voids in each dimension. For example, $\beta_{0}$ counts the number of connected components, $\beta_{1}$ is the number of one-dimensional loops and $\beta_{2}$ counts the number of enclosed spaces or voids. 
In this study, we used persistent homology to generate Betti curves independently associated with GM, WM, and CSF. We hypothesized that the features derived from the resulting Betti curves can serve as accurate predictors of chronological age. 

To compute the Betti curves of MRI scans, this study uses \textit{pixel level filtration} function on GM, WM and CSF segmented MRI volumes. Using pixel-level filtration we analyze the intensity values of the MRI pixels to study the topology of these tissues at different scales. Performing pixel-level filtration on GM, WM, and CSF images to compute Betti curves using persistent homology offers the advantage of capturing fine-grained topological features directly from raw imaging data, enhancing sensitivity to subtle structural changes across different tissue types. For example, pixel-level filtration allows for the detection of subtle structural variations, such as cortical thinning or irregularities in GM. For WM, filtration at the pixel level can reveal microstructural lesions or disconnections, for identifying demyelination, small lesions, or other WM abnormalities. For CSF, pixel-level filtration can detect minor changes in fluid distribution, such as small shifts in ventricle size or the presence of abnormal fluid accumulations. Thus, by analyzing topological features at multiple scales, pixel-level filtration enables the detection of minor abnormalities that could indicate early signs of pathological aging, such as those observed in MCI.
The Betti curves are computed directly on the 3D MRI volumes, and the filtration proceeds over voxel intensity values, effectively constructing a \textit{cubical complex} at each threshold level. As the filtration threshold increases, the cubical complex captures the appearance and disappearance of topological features, such as connected components ($\beta_{0}$) and loops ($\beta_{1}$), across the entire volume. This approach preserves the full 3D spatial structure of the tissue, unlike slice-wise 2D methods, and allows persistent homology to detect subtle 3D structural variations that may indicate early pathological changes. The voxel-level resolution ensures that no structural detail is lost during the filtration process, maximizing sensitivity to subtle yet clinically relevant anatomical changes associated with MCI.

\begin{figure*}
\centerline{\includegraphics[width=18cm]{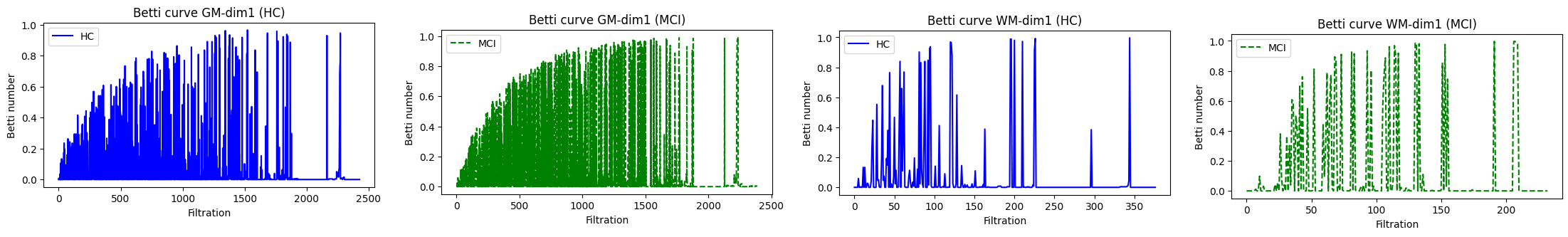}}
\caption{Betti curves generated from GM (\textit{col-1 and 2}) and WM (\textit{col-3 and 4}) segmentation using persistent homology for one representative HC (\textit{blue}) and MCI (\textit{green}) subject. These curves illustrate the evolution of topological features (e.g. dim-1 or loops) across different filtration scales, capturing structural differences in brain tissues between healthy and cognitively impaired subjects.
} 
\label{fig2}
\end{figure*}

In our analysis, we extract a range of both time-domain and frequency-domain features from the Betti curves corresponding to Betti dimension-1 and -2 for brain age prediction and disease classification. The time-domain features include statistical measures such as skewness, kurtosis, mean, variance, energy, and mean absolute deviation. These features provide insights into the distribution and variability of the Betti curves over time, capturing important characteristics of the topological structures within the data.
In addition to time-domain features, we also derive frequency-domain features to analyze the spectral properties of the Betti curves. These features include peak frequency, power spectral density, spectral entropy, spectral centroid, spectral roll-off, spectral flatness, and spectral flux.

\section{Experimental Results and Analysis}
The extracted comprehensive sets of features, both time-domain and frequency-domain, are used as inputs for machine learning models for brain age prediction and classification.

\subsection{Brain Age Prediction}
\begin{figure}
\centerline{\includegraphics[width=8cm]{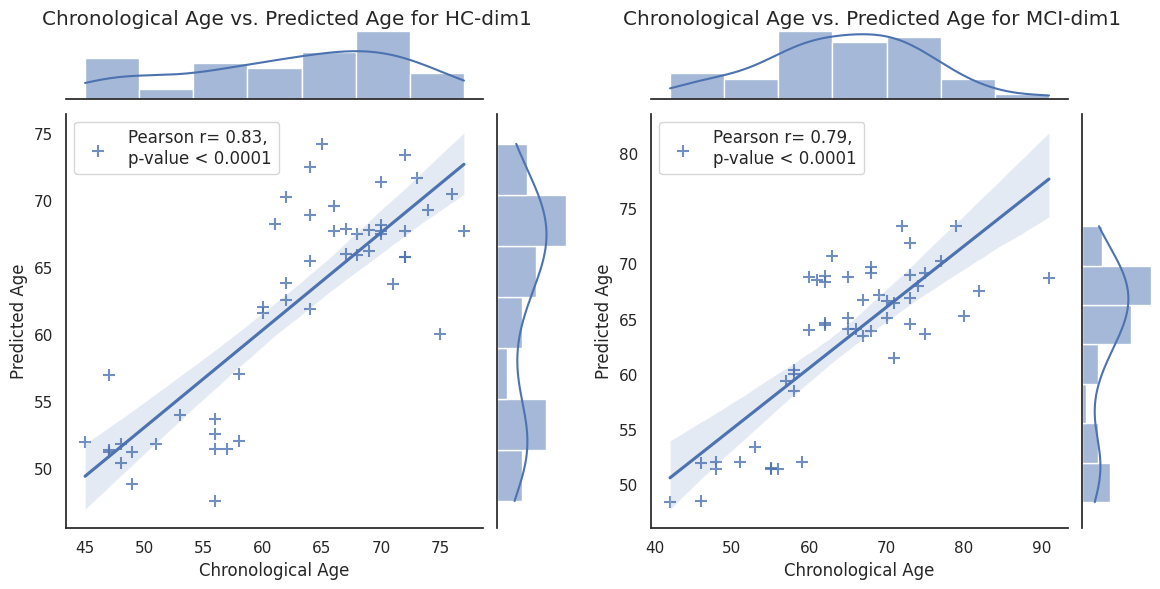}}
\caption{Correlation plot comparing the predicted brain age using features derived from Betti curves against the actual chronological age for both HC (\textit{left}) and MCI (\textit{right}) subjects.} 
\label{fig3}
\end{figure}

\begin{figure}
\centerline{\includegraphics[width=9cm]{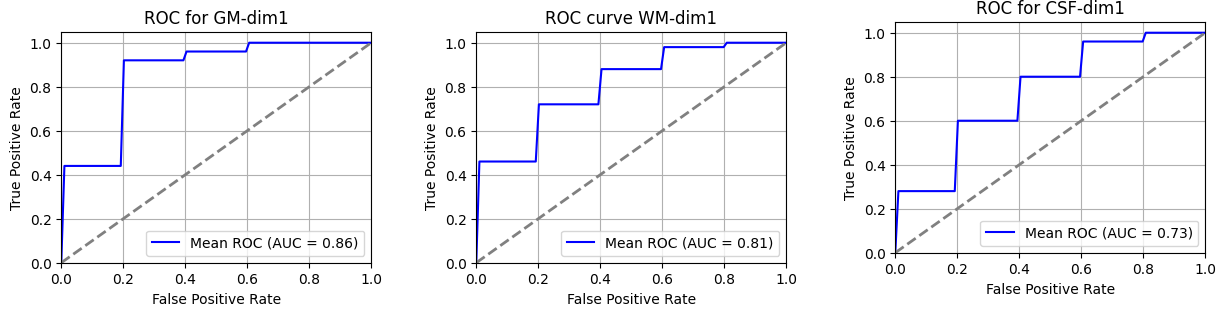}}
\caption {Receiver Operating Characteristic (ROC) curves showing the performance of SVM-based classification between healthy controls (HC) and individuals with mild cognitive impairment (MCI) using dimension-1 Betti curve features from different brain tissue regions GM, WM, CSF.} 
\label{fig4}
\end{figure}

Our initial experiments explore the potential of Betti curves as predictors of brain age. Figure~\ref{fig2} shows the Betti curves for GM and WM for one representative subject of HC and MCI. For brain age prediction, random forest regression model is employed. Figure~\ref{fig3} displays the correlation between choronological age and predicted age. Table~\ref{tab1} provides the mean absolute error (MAE) for HC and MCI subjects across different homology dimensions, with higher MAE values observed in MCI compared to HC. Specifically, the MAE difference for dimension-1 is 1.6 for GM and 1.53 for WM, while for dimension-2, the differences are 1.09 for GM and 0.92 for WM. The most significant differences in MAE are found in CSF homology features for both dimensions.
A higher brain age predictions for MCI individuals as compared to HC indicate that the model is capturing accelerated age-related changes, which can serve as a marker of cognitive decline or an early sign of neurodegeneration. The Betti curve features are also employed for classification tasks using a support vector machine with recursive feature elimination. Stratified 5-fold cross-validation with an 80:20 train–test split is used to ensure balanced representation of MCI and HC subjects across folds. Among all Betti dimensions, features derived from the $\beta_{1}$ Betti curves yields the highest classification accuracy of 80\%. This result is summarized in Table~\ref{tab2}. 
The receiver-operating-characteristics (ROC) curve is shown in Figure~\ref{fig4}.

\begin{table}
\caption{MAE for HC and MCI for brain age prediction}
\label{tab1}
\centering
\begin{tabular}{|l|l|l|l|l|}
\hline
Dim. & Region & MAE-HC &MAE-MCI &$\Delta$MAE\\
 & & (in years) &(in years) & (in years)\\
\hline
1 &GM & 3.9 & 5.5 &1.6\\
\cline{2-5}
 &WM &3.91 &5.44 &1.53\\
 \cline{2-5}
 &CSF &3.6 &5.78 &2.18\\
 \hline
2 &GM & 4.01 & 5.1 &1.09\\
\cline{2-5}
 &WM &4.6 &5.52 &0.92\\
 \cline{2-5}
 &CSF &4.1 &5.55 &1.45\\
 \hline 
\end{tabular}
\end{table}

\subsection{Normal Aging Vs. Pathological Aging}

\begin{figure*}
\centerline{\includegraphics[width=15cm]{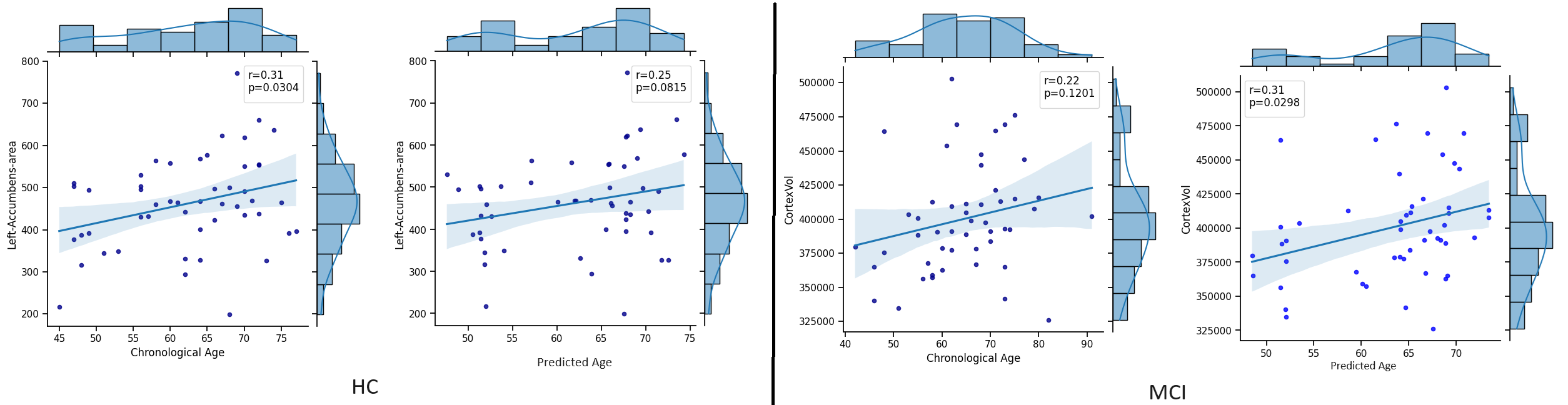}}
\caption{Visualization of differences in clinical imaging features between healthy aging and pathological aging (MCI). 
The \textit{left-side column} illustrates clinical features that correlate significantly with chronological age but not with predicted brain age, representing typical structural changes due to healthy aging. The \textit{right-side column} displays features significantly correlated with predicted brain age but not with chronological age in MCI subjects, highlighting early structural abnormalities indicative of pathological aging and cognitive decline.
} 
\label{fig5}
\end{figure*}

To differentiate between normal and pathological aging, we analyzed the correlation between clinical features, including FreeSurfer volumetric measures, area, thickness, and mean curvature, predicted brain age and chronological age. This analysis aimed to identify clinical features that yield statistically significant correlations ($p<0.05$). The findings are categorized into the following three distinct cases:

\textit{Case 1: Clinical features significantly correlated with predicted brain age but not with chronological age:} 
For individuals with MCI, significant correlations with predicted brain age suggest that these features capture structural alterations associated with cognitive decline or pathological aging processes. Such correlations may indicate accelerated brain aging or neurodegeneration related to the disease. Features exhibiting these correlations could serve as valuable biomarkers for detecting and monitoring cognitive impairment in MCI, offering insights into disease progression and response to treatment. Examples of these volumetric features include BrainSegVol, BrainSegVolNotVent, lhCortexVol, rhCortexVol, CortexVol, TotalGrayVol, SupraTentorialVol, and SupraTentorialVolNotVent. 
The right-side column of Figure~\ref{fig5} illustrates one example of this case for MCI subjects. 
In HC, these features may reflect subtle structural changes that are not strictly related to chronological age but rather to variations in brain health or integrity. The observed correlations might indicate early signs of age-related changes or individual differences in brain structure that have not yet impacted chronological age. Such features could serve as sensitive biomarkers for detecting deviations from typical brain aging in healthy individuals. In HC, examples of such volumetric features include Brainstem, CSF, and Right-Cerebellum-White-Matter.

\textit{Case 2: Clinical features not significantly correlated with predicted brain age but correlated with chronological age:}
For features that do not show significant correlations with predicted brain age but do correlate with chronological age suggests that these features represent typical age-related changes in brain structure. These features reflect established patterns of normal aging that are observed in HC and are not necessarily indicative of pathological processes. Examples of such features in HC include the Left-Accumbens-area, Right-Caudate, and CC posterior in HC. Notably, in our study, we did not find any such features in case of MCI. 
The left-side column of Figure~\ref{fig5} illustrates one example of this case. 

\textit{Case 3: Clinical features significantly correlated with both predicted brain age and chronological age:}
Features that exhibit significant correlations with both predicted brain age and chronological age serve as general markers of brain health and aging-related changes. These features likely reflect underlying mechanisms of both normal aging and cognitive decline. The observed correlations suggest that these features capture typical age-related changes and may also capture early signs of pathological alterations in brain structure. Examples of these features include Right-Accumbens-area, BrainSegVol-to-eTIV, and EstimatedTotalIntraCranialVol in HC. Similar to Case-2, in this case also we did not find any such features for MCI.

\subsection{Comparison with state-of-the-art baseline study}
We compared our proposed Betti-curve framework with the baseline approach reported by Saadat-Yazdi et.al. \cite{ref2}, demonstrated that Betti-curve topological features are strong predictors for brain age prediction of Alzheimer’s disease. Notably, our brain age estimation for MCI subjects achieved MAE that is comparable to the MAE reported by Saadat-Yazdi et.al. (MAE$\approx$5.5 years). This suggest that the proposed approach captures age-related structural changes effectively even in MCI cohort. 
Furthermore, for MCI classification, our framework achieves 80\% accuracy, surpassing the accuarcy of 74.4\% as reported by the baseline method that used residual Betti curves\cite{ref2}. These results highlight that incorporating both time-domain and frequency-domain features from Betti curves provides a measurable performance lift over existing approach as reported by Saadat-Yazdi et.al.\cite{ref2}.

\subsection{Clinical Relevance of the Study}
This study demonstrates significant clinical relevance by introducing a novel, non-invasive framework for the early detection and monitoring of cognitive decline, particularly MCI, using sMRI and persistent homology-based topological features. The discrepancy between predicted brain age and chronological age provides an intuitive and quantifiable marker for assessing accelerated brain aging, which is often associated with MCI and dementia. Clinicians could use this measure to flag at-risk individuals in a routine scan.
Moreover, the present study examines to distinguish between features of normal aging and those of pathological aging, aiding neurologists in identifying whether observed brain changes are disease-related or part of healthy aging. This distinction is critical for avoiding misdiagnosis and unnecessary treatment. By identifying clinical features that correlate specifically with predicted brain age in MCI patients, the method allows for personalized monitoring of disease progression. 

However, the study is conducted on a relatively small dataset comprising only 100 sMRI scans (50 MCI and 50 healthy controls), limiting the generalizability of the findings. Future studies should incorporate larger, multi-center datasets with more diverse demographic and clinical profiles to improve the statistical power and ensure broader applicability of the results. Also, extending the framework to longitudinal MRI data will enable the modeling of temporal changes in brain topology, could provide deeper insights into disease progression and the impact of therapeutic interventions.

\begin{table}
\caption{Classification using dimension-1 Betti features}
\label{tab2}
\centering
\begin{tabular}{|l|l|l|l|l|l|l|}
\hline
 & Sub. & Precision &Recall &F1- &Accuracy\\
 & &  & &score &\\
\hline
GM- &HC &0.78 &0.84 & 0.81 & 80\% \\
\cline{2-5}
dim1 &MCI &0.83 &0.76 &0.79 & \\
\hline
WM- &HC &0.74 &0.92 & 0.82 & 80\% \\
\cline{2-5}
dim1 &MCI &0.89 &0.68 &0.77 & \\
\hline
CSF- &HC &0.70 &0.84 & 0.76 & 74\% \\
\cline{2-5}
dim1 &MCI &0.80 &0.64 &0.71 & \\
\hline
\end{tabular}
\end{table}

\section{Conclusion}

The present study explores the use of Betti curves features- especially those from dimension-1 and 2 serve as effective biomarkers for capturing structural brain changes associated with aging. Moreover, three distinct categories of clinical features are identified based on their significant correlation with predicted brain age and chronological age, highlighting the potential of Betti curves for differentiating normal aging from pathological aging.

\bibliographystyle{IEEEbib}
\bibliography{strings,refs}

\end{document}